\documentstyle[epsfig]{aipproc}
\begin{document}

\title{MOND in the Early Universe}
\author{Stacy McGaugh}
\address{Department of Astronomy, University of Maryland, College Park, MD
	20742}

\maketitle

\begin{abstract}
I explore some consequences of Milgrom's modified dynamics for cosmology.
There appear to be two promising tests for distinguishing MOND from CDM:
(1) the rate of growth of structure
and (2) the baryon fraction.  These should be testable with observations
of clusters at high redshift and the microwave background, respectively.
\end{abstract}

\section*{Standard Cosmology}

The standard hot big bang cosmology has many successes; too many to list here.
The amount of data constraining cosmic parameters has increased rapidly,
until only a small region of parameter space remains viable.  This has led
to talk of a `concordant' cosmology with $\Omega_{\cal M} \approx 0.3$ and
$\Omega_{\Lambda} \approx 0.7$ \cite{OS}.

This is a rather strange place to end up.  The data do not {\it favor\/} these
parameters so much as they {\it disfavor\/} other combinations more.
A skeptic might suspect that concordance is merely the
corner we've painted ourselves into prior to the final brush stroke.

This is not an idle concern, as there remains one major outstanding problem:
dark matter.  Something like 90\% of the universe is supposedly made of stuff
we can not see.
There are, to my mind, two ironclad lines of reasoning that
{\it require\/} the dark matter to be nonbaryonic, cold dark
matter (CDM) like WIMPs or axions.
\begin{enumerate}
\item $\Omega_{\cal M} \gg \Omega_b.$
\item Structure does not have time to grow from a smooth microwave
	background to the rich structure observed today
	without a mass component whose perturbations can
	grow early without leaving an imprint on the CMBR.
\end{enumerate}
{\bf BUT} we have yet to detect WIMPs or axions.
Their existence remains an unproven, if well motivated, assumption.

\section*{Is there any Dark Matter?}

It is often stated that the evidence for dark matter is overwhelming.
This is not quite correct: the evidence for {\it mass discrepancies\/}
is overwhelming.  These might be attributed to {\it either\/} dark matter
{\it or\/} a modification of gravity.

Rotation curves played a key role in establishing the mass discrepancy problem,
and remain the best illustration thereof.
There are many fine-tuning problems in using dark matter to explain
these data.  I had hoped that the resolution of these problems would
become clear with the acquisition of new data for low surface brightness
galaxies.  Instead, the problems have become much worse \cite{DM}.

These recent data are a particular problem for CDM models, which simply do not
fit \cite{me99}. Tweaking the cosmic parameters can reduce but
not eliminate the problems.  No model in the concordant range can fit the
data unless one invokes some {\it deus ex machina\/} to make it so.

There is one theory which not only fits the recent observations,
but predicted them \cite{MD}.
This is the modified dynamics (MOND) hypothesized by Milgrom \cite{Mb}.
The basic idea here is that instead of dark matter, the force law is
modified on a small acceleration scale, $a_0 \approx 1.2
\>{\rm \AA}\,{\rm s}^{-2}$. For $a \gg a_0$ everything is normal, but for
$a \ll a_0$ the effective force law is $a = \sqrt{a_N a_0}$, where $a_N$ is
the usual Newtonian acceleration.

This hypothesis might seem radical, but it has enormous success in
rectifying the mass discrepancy.  It works exquisitely well in rotating
disks where there are few assumptions \cite{BBS,S96,SV,mondLSB}.
MOND also seems to work in other places, like dwarf Spheroidals
\cite{MD,GS,7dw,MOVK}, galaxy groups \cite{groups},
and filaments \cite{filam}.  The only place in which it
does not appear to completely remedy the mass discrepancy is in the cores of
rich clusters of galaxies \cite{clust2}, a very limited missing mass problem.

It is a real possibility is that MOND is correct, and CDM does not exist.
Let us examine the cosmological consequences of this.

\section*{Simple MOND Cosmology}

There exists no complete, relativistic theory encompassing MOND, so a
strictly proper cosmology can not be derived.  However, it is possible
to obtain a number of heuristic results in the spirit of MOND.
The simplest approach is to assume that $a_0$
does not vary with cosmic time.  This need not be the case
\cite{CoA,AnnPh}, but makes a good starting point.  I do not have space to 
derive anything here, and refer the reader to detailed published work
\cite{CoA,AnnPh,Scosm}.

Making this simple assumption,
the first thing we encounter is that it is not trivial to
derive the expansion history of the universe in MOND \cite{Scosm,F84}.
This might seem unappealing, but does have advantages.  For example,
a simple MOND universe will eventually recollapse
irrespective of the value of $\Omega_{\cal M}$.
There is no special value of $\Omega_{\cal M}$, so no flatness problem.

Conventional estimates of $\Omega_{\cal M}$ are overly large in MOND.
Instead of $0.2 < \Omega_{\cal M} < 0.4$,
MOND gives $0.01 < \Omega_{\cal M} < 0.04$.
So a MOND universe is very low density, consistent with being composed
purely of baryons in the amount required by big bang nucleosynthesis.

This makes some sense.  Accelerations in the early
universe are too high for MOND to matter.
This persists through nucleosynthesis and recombination,
so everything is normal then and all the usual results are retained.
MOND does not appear to contradict any empirically established cosmological
constraint.

The universe as a whole transitions into the MOND regime ($c H_0 \sim a_0$)
around $z \sim 3$, depending on $\Omega_{\cal M}$.
Sub-horizon scale bubbles could begin to
make this phase transition earlier, providing seeds for the growth of
structure and setting the mass scale for galaxies \cite{Scosm}.
Nothing can happen until the radiation releases its grip on the
baryons ($z \sim 200$), by which time the typical acceleration is quite
small.  As a result, things subsequently behave
as if there were a {\bf lot} of dark matter: perturbations grow fast.
This provides a mechanism by which structure grows from a smooth state to
a very clumpy one rapidly, without CDM.  

Now recall the two ironclad reasons why we must have CDM.  In the case of
MOND
\begin{enumerate}
\item $\Omega_{\cal M} = \Omega_b \approx 0.02$
\item There is no problem growing structure rapidly from a smooth CMBR
	to the rich amount seen at $z = 0$ with baryons alone.
\end{enumerate}

\section*{Predictions}

The simple MOND scenario makes two predictions which distinguish it from
standard CDM models.
\begin{enumerate}
\item Structure grows rapidly and to large scales.
\item The universe is made of baryons.
\end{enumerate}

There are indications that at least some galaxies form early, and
are already clustered at $z \approx 3$ \cite{GSADPK}.
At low redshifts, we are continually surprised by the size of the largest
cosmic structures.  It makes no sense in the conventional context that
fractal analyses should work as well as they do \cite{fractal}.
A MOND universe need not be precisely fractal, but if analyzed in this way
it naturally produces the observed dimensionality \cite{Scosm}.
So there are already a number of hints of MOND-induced behavior in cosmic
data.

A strong test may occur for rich clusters.
These are rare at $z > 1$ in any CDM cosmology \cite{ECF}.
In the simple MOND universe, clusters form at $z \approx 3$ \cite{Scosm}.
Upcoming X-ray missions should be able to detect
these \cite{Mush}.

\begin{figure}[tk.eps]
\centerline{\epsfig{file=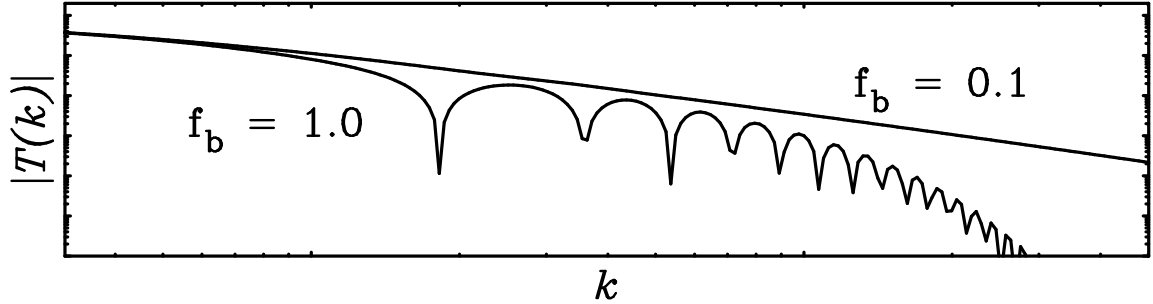,width=6in}}
\caption{Post-recombination transfer functions for low and
high baryon fraction models.}
\label{fig1}
\end{figure}

The rapid growth of perturbations in MOND overcomes the usual objections
to purely baryonic cosmologies.  The baryon fraction
makes a tremendous difference to the bumps
and wiggles in the CMBR power spectrum (Figure \ref{fig1}) \cite{EH}.
CDM smooths out the acoustic oscillations
due to baryons in a way which can not happen if $f_b =1$.

This should leave some distinctive feature in the CMBR that can be
measured by upcoming missions like MAP \cite{Sperg}.  Unfortunately, it is
easy only to predict the spectrum as it emerges shortly after recombination.
Since the growth of structure is rapid and nonlinear
in MOND, there might be a strong integrated Sachs-Wolfe effect.
I would expect this to erase any hint of the oscillations
in the $z=0$ galaxy distribution, but not necessarily in the CMBR.
The initial spectrum in the CMBR is
sufficiently different in the CDM and MOND cases that there is a good prospect
of distinguishing between the two.

\end{document}